\documentclass[a4paper,fleqn,usenatbib]{mnras}

\usepackage{newtxtext,newtxmath}

\usepackage[T1]{fontenc}
\usepackage{ae,aecompl}

\usepackage{graphicx}	
\usepackage{bm}
\usepackage[ruled]{algorithm2e}
\usepackage{color}

\newcommand*{\obs}{\ensuremath{\mathbf{y}}\xspace}
\newcommand*{\params}{\ensuremath{\boldsymbol{\theta}}\xspace}
\newcommand*{\Lparams}{\ensuremath{\boldsymbol{\lambda}}\xspace}

\newcommand*{\post}{\ensuremath{p(\boldsymbol{\theta}|\mathbf{y})}\xspace}
\newcommand*{\prior}{\ensuremath{p(\boldsymbol{\theta})}\xspace}
\newcommand*{\lh}{\ensuremath{p({\mathbf{y}}|\boldsymbol{\theta})}\xspace}
\newcommand*{\marginal}{\ensuremath{p(\mathbf{y})}\xspace}
\newcommand*{\target}{\ensuremath{\pi(\boldsymbol{\theta}|\mathbf{y})}\xspace}

\newcommand*{\noise}{\mathbf{e}\xspace}

\newcommand{\mbE}{\mathbb{E}}

\title[Adaptive Target Adaptive Importance Sampling]{A Bayesian inference and model selection algorithm with an optimisation scheme to infer the model noise power}

\author[J. L\'opez-Santiago et al.]{
J. L\'opez-Santiago,$^{1}$\thanks{E-mail: jalopezs@ing.uc3m.es}
L. Martino,$^{3}$
M. A. V\'azquez$^{1}$
and J. Miguez$^{1}$
\\
$^{1}$Dep. of Signal Theory and Communications, 
                  Universidad Carlos III de Madrid, 
                  Av. de la Universidad 30, E-28911, 
                  Legan\'es, Spain\\
$^{3}$Dep. of Signal Theory and Communications, 
                  Telematic Systems and Computation, 
                  Universidad Rey Juan Carlos, 
                  Camino del Molino s/n. E-28943, Fuenlabrada, Spain
}

\date{Accepted XXX. Received YYY; in original form ZZZ}

\pubyear{2015}

\begin{document}
\label{firstpage}
\pagerange{\pageref{firstpage}--\pageref{lastpage}}
\maketitle

\begin{abstract}
Model fitting is possibly the most extended problem in {science}. {Classical approaches include  the use of} least-squares fitting procedures and maximum likelihood methods to {estimate} the value of the parameters {in} the model. {However}, in recent years, Bayesian inference tools {have gained traction}. Usually, Markov chain Monte Carlo methods are applied to inference problems, but they present some disadvantages, particularly when comparing different models {fitted} {to} the same dataset. Other Bayesian methods can deal with this issue {in a natural and effective way}. 
We have implemented an importance sampling algorithm adapted to {Bayesian inference problems in which the power of the noise in the observations is not known a priori}. The main advantage of importance sampling is that the {model evidence} can be derived directly from the so-called importance weights {-- while MCMC methods demand considerable postprocessing.}
{The use of our {adaptive target, adaptive importance sampling (\emph{ATAIS})} method is shown by inferring, {on the one hand,} the parameters of a simulated flaring event which includes a damped oscillation {and, on the other hand, real data from the \emph{Kepler} mission}.} ATAIS includes a novel {automatic adaptation of the target distribution}. It {automatically} {estimates} the variance of the {noise} {in the model}. ATAIS admits parallelisation, {which} decreases {the computational run-times} notably.
{We compare our method against a nested sampling method within a model selection problem.}

\end{abstract}




\section{Introduction}
\label{sec:Intro}

{{Bayesian data analysis has become a popular tool in many research fields}. A {particular} {type of} problem in which Bayesian analysis is {often} used {consists in} fitting models to empirical data {in order to} {estimate} the value of {their unknown parameters}. {This is} referred to as {Bayesian inference}. Two different methods are commonly applied in Bayesian inference: Markov chain Monte Carlo (MCMC) and importance sampling (IS). In the Astrophysical literature, it is {common} to find applications of various MCMC algorithms {\citep[see][for some recent examples]{Caruso2019, Komanduri2020, Martinez2020}.} Contrarily, IS methods are {rarely} used {to infer parameters \citep{Wraith2009, Lewis2011}, but to determine the Bayesian evidence with the accepted samples from an MCMC-based method \citep[e.g.][]{Perrakis2014, Nelson2016}}.

One reason for the {popularity} of MCMC methods is that they are easy to implement. However, MCMC methods present some {significant} disadvantages. A well-known problem is that they generate correlated samples \citep{Martino2017b}, although {different} methods may be applied to {mitigate this problem} {\citep[e.g.][]{Pascoe2020}}. In addition, MCMC algorithms are difficult to parallelise {(although different chains can be run in parallel)}. However, the main disadvantage of MCMC is its difficulty for determining the {model evidence} {(a.k.a. marginal likelihood)} directly and, hence, comparing different models for the same dataset. 
{A solution for that problem is proposed by \citet{Green1995}, where the author presents a method to create reversible Markov chain samplers that jump between models with different dimensionality.}
In recent years, MCMC methods {have} been applied extensively in exoplanet searches. Some examples are \citet{Gregory2011}, \citet{Barros2016}, \citet{Affer2019}, and \citet{Trifonov2019}, among many others. A list of {research} groups working on this issue can be found in \citet{Dumusque2017} {and in \citet{Nelson2020}}. MCMC are also used in spectroscopic studies \citep{Greene2018,Casasayas2020}. 

{Compared to MCMC methods, IS techniques offer a simpler and much more natural solution to the problem of model selection, as they readily yield estimates of the model evidence without additional computations. Moreover, IS algorithms admit {simple} parallelisation and they can be combined with MCMC methods to permit larger coverage of the state space \citep{Martino2018}.}
{Besides,} ``the IS methods are elegant, theoretically sound, simple-to-understand, and widely applicable'' \citep{Bugallo2017}.  {Several of these methods are} available in the specialised literature \citep{Cappe2004,Cornuet2012,Elvira2015} and most of them involve some kind of adaptation {for} the mean and variance of the proposal distribution \citep{Martino2017}. {Like} MCMC, IS methods admit tempering or annealing schemes \citep{Swendsen1986}. {The power of the IS methods resides in the capability of inferring the model parameters together with the marginal likelihood.} 

{Distinct methods to compute the marginal likelihood are available in the literature \citep[see][for an extensive review]{Llorente2021}. These can be classified into four families. A first class of methods are those based on deterministic approximations and density estimation, like the Bayesian-Schwarz information criterium \citep[BIC;][]{Schwarz1978}. Under some, sometimes weak assumption, these methods approach the Bayes factor or they estimate the value of the posterior density, around previously chosen samples. The samples are obtained using some sampling method like acceptance-rejection (ARM), MCMC or others. The second family corresponds to the techniques based on IS. They draw samples from one or more proposal distribution and determine weights for them. These weights can be used to approximate the marginal posterior directly without additional computations \citep{Elvira2017}. A third class of methods are those based on the joint use of MCMC and IS. They combine the capability of MCMC methods to explore the parameter/state space and that of IS to approximate the marginal posterior \citep{Perrakis2014,Nelson2016,Martino2018}. The last family of methods are based on quadrature schemes. They are settled on the Lebesgue representation of the integral of the marginal likelihood \citep{Llorente2021}. The most often used of them is the so-called nested sampling \citep[see][for a complete review]{Buchner2021a}. Nested sampling (NS) is a method developed specifically for the problem of model selection \citep{Skilling2006}. Nevertheless, currently most available codes incorporate a bayesian inference algorithm \citep[][]{Speagle2020,Buchner2021b}.
In practice, NS estimates the marginal likelihood by quadratures through the use of nodes. These nodes are particular samples drawn with a sampling method, commonly MCMC \citep{Feroz2019}.}

{A good example of a model selection problem in astrophysics is exoplanet detection through radial velocity curves. Inference must be performed on the five parameters defining the planet's orbit. In addition, a sixth parameter is needed to account for the mean radial velocity of the star. For each planet additionally included in the model, another five parameters need to be inferred. Hence, the number of dimensions grows rapidly and models must be compared by means of the marginal likelihoods. In addition, the {statistics of the observational noise are not known a priori and inference on them is not easy. 
%
Several attempts have been made in the past to introduce IS for exoplanet search. \citet{Loredo2011} developed an adaptive IS (AIS) method to estimate the marginal likelihood to compare between models with planets and without planets. However, that is still an open line of research. Previously, \citet{Hogg2010} had applied IS to infer the distribution of the {planet's} eccentricity. The authors indicate that the method can be applied to any other model parameter. Finally, \citet{Liu2014} proposed an AIS method with annealing to explore the marginal posterior pdfs. In any case, the problem of {estimating} the actual power of the {noise affecting the observations} is not tackled. 

{Another example of a process with many parameters to infer is flaring events that trigger plasma oscillation \citep[e.g][]{Mathioudakis2006,Nakariakov2007,stepanov2012,Reale2016,LS2016,LS2018,Nakariakov2020}. Here, the distinct models represent different decay functions of the oscillation and the flare intensity in the observed light curve \citep{Pascoe2020}. A minimum of eight parameters are needed for an exponentially damped oscillation with constant background. The number of model parameters increases rapidly if a non uniform background or multi exponential decay are included.}

{In this work, we {present an implementation of} an {IS} scheme {for model inference} that includes {adaptation and weight clipping} {\citep{Koblents2014}}.  {We also incorporate a novel target adaptation procedure}. The method is {termed ATAIS}. {A formal statistical study of the adaptation procedure, including its performance and comparison with actual marginal posteriors from a multimodal distribution was carried out in \citet{Martino2021}.} The details of the method together with theoretical aspects are explained in the next sections. {They include several algorithms to implement the method in the reader preferred programming language.} The main advantage of ATAIS with respect to other AIS methods is that it incorporates an optimisation procedure to determine the power of the observation noise automatically. {This optimisation scheme avoids the use of annealing and the need to infer the variance of the noise together with the remaining model parameters. After running of ATAIS, the posterior distribution of the noise variance is obtained as explained in Section~5 of \citet{Martino2021}. Furthermore, a global bayesian evidence can be determined that includes the noise variance as a parameter instead of fixing it.}
}

This article is structured as follows. {Bayesian model fitting is briefly introduced} in Section~\ref{sec:bayesFit}. Section~\ref{sec:IS} shows a formal, brief description of the  {IS} method. Its application to Bayesian inference problems is {described} in Section~\ref{sec:BayesianInference}, together with a generic algorithm. Section~\ref{sec:ATAIS} describes {the new} ATAIS algorithm, an IS scheme {with adaptation of the proposal distribution including a procedure to infer the variance of the noise}. {Section~\ref{sec:ModelSelection} presents an application of the method for model selection. The accuracy of the results is compared with that of nested sampling methods.} {An application of the method to detect and characterise a damped oscillation in a flare light curve} is presented in Section~\ref{sec:Simul}. Conclusions are given in Section~\ref{sec:conclusions}.

\section{Bayesian approach to model fitting}
\label{sec:bayesFit}

 {Consider a physical process {described} by a {certain} model. The latter {depends on one or more} {unknown parameters $\boldsymbol{\lambda} \in \mathbb{R}^{d_\lambda}$} {that have to be} inferred by fitting the model to empirical data $\obs \in \mathbb{R}^{d_y}$. Commonly, observations are affected by noise, {which we assume additive}. Therefore, {we can write}
\begin{equation}
\mathbf{y} = {\bf f} (\boldsymbol{\lambda}) + \noise,
\label{eq:model}
\end{equation}
where $\noise \in \mathbb{R}^{d_y}$ is a random vector that follows a probability distribution with zero mean and {covariance matrix $\sigma^2 \mathbf{I}_{d_y}$ ($\mathbf{I}_{d_y}$ denotes the $d_y\times d_y$ identity matrix)} and $\mathbf{f} : \mathbb{R}^{d_\lambda} \rightarrow \mathbb{R}^{d_y}$ is the {transformation that maps the model parameters into the space of the observations}. Usually, {the noise power parameter, {hereby denoted by} $\sigma^2$} is not known and must be inferred too. Let $\params$ be the set of parameters to be inferred, which includes the noise power, i.e. 
\begin{equation}
\params = [\lambda_1, \lambda_2, ..., \lambda_n, \sigma^2].
\label{eq:AllParams}
\end{equation}
where {$n+1$} is the number of parameters in the model.
}

{From a Bayesian perspective, $\params \in \mathbb{R}^{d_{\theta}}$ is a set of random variables. Its} {posterior probability density function (pdf) given the data can be written by} applying the Bayes' theorem, {as}
\begin{equation}
\post = \frac{\lh \prior}{\marginal}  {\propto \lh \prior = \target},
\label{eq:bayes}
\end{equation}
%
where $\lh$ is the so-called likelihood function (the {pdf of the data conditional on} the set of parameters $\params$), $\prior$ is the prior distribution of $\params$ ({encoding the prior knowledge about the parameters}) and $\marginal$ is the model (or Bayesian) evidence\footnote{We use an argument-wise notation for pdfs where, given two random variables $x$ and $y$, $p(x)$ and $p(y)$ are the pdfs of $x$ and $y$, respectively, and they are possibly different functions. Moreover, $p(x,y)$ denotes the join pdf of the pair $(x,y)$ and $p(x|y)$ is the conditional pdf of $x$ given $y$.}. 
{Note that $\target$ is an unnormalised posterior function, i.e., a function proportional to $\post$. Generally, in real world applications, we are able to evaluate {(point-wise)} the function $\target$ instead of $\post$ because the normalising constant 
\begin{equation}
\marginal =\int_{\boldsymbol{\Theta}} \target d\boldsymbol{\theta} = \int_{\boldsymbol{\Theta}} \lh \prior d\boldsymbol{\theta}
\label{eq:marginalL0}
\end{equation}
cannot be computed.}
The minimum mean square error (MMSE) estimator of $\params$ is \citep{Pearson1894}
\begin{equation}
\boldsymbol{\theta}_{\mathrm{MMSE}}  = \int_{\boldsymbol{\Theta}} \params \post d\params,
\label{eq:MMSE}
\end{equation}
where $\boldsymbol{\Theta} \subseteq \mathbb{R}^{d_{\params}}$ is the parameter space. {To be specific, this estimator coincides with the conditional expectation of the random parameter vector $\boldsymbol{\theta}$ given the available observations $\mathbf{y}$ --hence, we denote $\boldsymbol{\theta}_{\mathrm{MMSE}} = \mbE[ \boldsymbol{\theta} | \mathbf{y} ]$.} {Other  {classic estimators} {which are} often used are the maximum a posteriori (MAP) estimator,
\begin{equation}
\boldsymbol{\theta}_{\mathrm{MAP}}=\arg \max_{\boldsymbol{\Theta} } \post,
\end{equation}
or the median of the posterior distribution, $\boldsymbol{\theta}_{\mathrm{MED}}$.  {None of} these point-wise estimators, $\boldsymbol{\theta}_{\mathrm{MMSE}}$, $\boldsymbol{\theta}_{\mathrm{MAP}}$, $\boldsymbol{\theta}_{\mathrm{MED}}$, {can} be computed analytically in most  {real world scenarios}. Furthermore, interval estimations are often required in order to provide uncertainty analysis and  {outlier detection}, for instance.
Approximations of the posterior distribution, $\post$, and related} integrals are required to compute all these quantities. 

One  {approach to the approximation of} the posterior distribution is the use of Monte Carlo (MC) algorithms. They provide a sample-based approximation of the posterior (with a random support) that can be obtained virtually for any kind of posterior {distribution} and any dimension of the inference space. The obtained population of samples can be employed  {to approximate} integrals involving the posterior distribution. One important family of MC methods are the so-called Markov Chain Monte Carlo algorithms. Another important family of MC methods are the Importance Sampling (IS) techniques. There are several advantages in using IS with respect to MCMC  {(as mentioned in Section~\ref{sec:Intro})}. One clear benefit of IS is the ability {to approximate} the model evidence $\marginal$ with an  {easy-to-implement estimator}. Estimation of the model evidence via MCMC requires much more  {sophisticated and, in general, less efficient methods}. 
{The main disadvantage is that they tend to be less efficient in exploring the parameter space.}

\section{Brief introduction to importance sampling {(IS)}}
\label{sec:IS}

Let us assume {that} we are interested in obtaining the expected value of a random variable $X$, {denoted} $\mathbb{E} [X]$. If the variable follows a distribution with pdf $p(x)$, its expected value is given by the integral
\begin{equation}
\mathbb{E}_p \left[ X \right] = \int_\mathcal{X} x p(x) dx
\label{eq:expected}
\end{equation}
The subindex $p$ indicates {that} the expectation is under $p(x)$. The integral {is over} the entire space of the variable, $\mathcal{X}$. For instance, if $X$ is  {a} unidimensional real variable, then $\mathcal{X} \equiv \mathbb{R}$. The integral in Eq.~\eqref{eq:expected} may not have {an} analytical solution. {The Monte Carlo approximation \citep{Metropolis1949} consists in drawing $N$ samples $x_1, x_2, ..., x_N$ from the density function} {$p(x)$ so that}
\begin{equation}
\mathbb{E}_p \left[ X \right] \approx \frac{1}{N} \sum\limits_{i = 1}^{N} x_i 
\label{eq:expectedMC}
\end{equation}
{Often, it is hard to draw samples from $p(x)$ directly. In such case, an auxiliary pdf, known as proposal function and denoted by $q(x)$, can be used to rewrite the expectation in Eq.~\eqref{eq:expected} as}
\begin{subequations}
\begin{align}
\mathbb{E}_p \left[ X \right] & =  \int_\mathcal{X} x p(x) dx  \\ 
                                            & =  \int_\mathcal{X} x \frac{p(x)}{q(x)} q(x) dx  \\ 
                                            & =  \int_\mathcal{X} x w(x) q(x) dx 
\end{align}
\label{eq:IS}
\end{subequations}
The last integral in Eq.~\eqref{eq:IS} is the expected value of the function $xw(x)$ under $q(x)$, {which we denote as $\mathbb{E}_q \left[ X w(X) \right]$. We have shown, therefore, that $\mathbb{E}_p \left[ X \right] =  \mathbb{E}_q \left[ X w(X) \right]$.} We can further elaborate on this expression to arrive at
%
\begin{subequations}
\begin{align}
\mathbb{E}_p \left[ X \right] & = \frac{\int_\mathcal{X} x w(x) q(x) dx}{\int_\mathcal{X} \frac{p(x)}{q(x)} q(x) dx}  \\ 
                                            & = \frac{\int_\mathcal{X} x w(x) q(x) dx}{\int_\mathcal{X} w(x) q(x) dx} = { \frac{ \mbE_q[ X w(X) ] }{ \mbE_q[ w(X) ] } }
\end{align}
\label{eq:ISNorm}                     
\end{subequations}
where it has been used that $p(x)$ is a pdf and, therefore, $\int_X p(x) dx = 1$. {From Eq.~\eqref{eq:ISNorm}, we readily obtain the} Monte Carlo approximation,
\begin{equation}
\mathbb{E}_p \left[ X \right] \approx \frac{\frac{1}{N} \sum\limits_{i = 1}^{N} x_i w(x_i)}{\frac{1}{N} \sum\limits_{i = 1}^{N} w(x_i)} = \sum\limits_{i = 1}^{N} x_i \bar{w}(x_i),
\label{eq:ISNormMC}  
\end{equation}
where 
\begin{equation}
\bar{w}(x_i) = \frac{w(x_i)}{\sum\limits_{i = 1}^{N} w(x_i)}
\label{eq:weights}  
\end{equation}
are normalised weights and {the $x_i$'s} are samples from $q(x)$. {The method is called} \emph{importance sampling}, and the $\bar{w}(x_i)$'s are  {referred to as} \emph{importance weights} {\citep{Robert2005}}. 

\section{Application to Bayesian inference}
\label{sec:BayesianInference}

{From Eq.~\ref{eq:expected}, the MMSE estimator of $\params$ given the data $\obs$ is the {conditional} expectation}
\begin{equation}
\mbE \left[ \params | \mathbf{y} \right]   = \int_{\boldsymbol{\Theta}} \params \post d\params.
\label{eq:expectedParams}
\end{equation}
Equation~\ref{eq:expectedMC} provides an approximation to this integral if enough samples are drawn from $\post$. However, the posterior distribution of $\params$ is usually {hard to sample}. Importance sampling {is simple to apply, though. Using the method outlined in Section~\ref{sec:IS}, we obtain}
%
\begin{subequations}
\begin{align}
\mathbb{E} \left[ \params | \mathbf{y} \right] & = \int_{\boldsymbol{\Theta}} \params \post d\params, \\
                                                       & = \int_{\boldsymbol{\Theta}} \params \frac{\lh \prior}{\marginal} d\params, \\
                                                       & = \frac{1}{\marginal} \int_{\boldsymbol{\Theta}} \params \target d\params, \\
                                                       & = \frac{1}{\marginal} \int_{\boldsymbol{\Theta}} \params \frac{\target}{q(\params)} q(\params) d\params, \\
                                                       & = \frac{1}{\marginal} \int_{\boldsymbol{\Theta}} \params w(\params) q(\params) d\params,
\end{align}
\label{eq:BayesianIS}
\end{subequations}
 {where $w(\params)=\frac{\target}{q(\params)}$.}
{The term $\target = \lh \prior \propto \post$} is usually referred to as the target function (or unnormalised posterior distribution). {A standard} Monte Carlo approximation {now yields}
\begin{equation}
\mathbb{E} \left[ \params | \mathbf{y} \right] \approx \frac{1}{N \marginal} \sum\limits_{i = 1}^{N} \params_i w(\params_i) 
 =: \widehat{I}_N, 
\label{eq:BayesianMonteCarloIS}
\end{equation}
where $\params_i\sim q(\params)$, $i=1, ..., N$, and $N$ is the number of samples. An alternative estimator (that avoids the need to compute $\marginal$) is obtained by normalising the weights $w(\params_i)$, namely
\begin{equation}
\mathbb{E} \left[ \params | \mathbf{y} \right] \approx \sum\limits_{i = 1}^{N} \params_i \bar{w}(\params_i) 
=:\widetilde{I}_N,
\label{eq:NormBayesianMonteCarloIS}
\end{equation}
where
\begin{equation}
{\bar{w}_i=\bar{w}(\params_i)} \approx \frac{w(\params_i)}{\sum\limits_{i = 1}^{N} w(\params_i)}
\label{eq:ParamsWeights}  
\end{equation}
{are normalised importance weights.}
Similarly, an estimate of the {covariance matrix is}
{
\begin{equation}
\widetilde{C}_N \approx \sum\limits_{i = 1}^{N} \left( \params_i - {\widetilde{I}_N} \right) \left( \params_i - {\widetilde{I}_N} \right)^\intercal \bar{w}_i, 
\label{eq:VarBayesianMonteCarloIS}
\end{equation}
}
Equations~\eqref{eq:NormBayesianMonteCarloIS}-\eqref{eq:VarBayesianMonteCarloIS} are used when the {model evidence} $\marginal$ is unknown and they provide a method for approximating the expected value of a parameter, or a set of parameters in the model, together with their covariance matrix {and any other statistics that may be of interest}. Note also that {estimates} of $\boldsymbol{\theta}_{\mathrm{MAP}}$ and $\boldsymbol{\theta}_{\mathrm{MED}}$ can be easily obtained {using the set} of weighted samples. 
For instance, $\boldsymbol{\theta}_{\mathrm{MAP}}$ is given by $\boldsymbol{\widehat{\theta}}_{\mathrm{MAP}}=\arg\underset{1 \le i \le N}{\max~} \pi(\params_i|{\bf y})$.

\subsection{Estimation of the model evidence}
\label{sec:Marginal}

A great advantage of using importance sampling is that the {model evidence} $\marginal$ can be {estimated} directly from the individual weights of each Monte Carlo sample ${w}(\params_i)$. As a result, a comparison between models can be  {easily done}  ({while} other methods, like MCMC, require post-processing {\citep[][]{Ford2007, Perrakis2014, Nelson2016, Pascoe2020}}. 
From Eq.~\eqref{eq:bayes},
\begin{equation}
\marginal =\int_{\boldsymbol{\Theta}} \target d\boldsymbol{\theta} = \int_{\boldsymbol{\Theta}} \lh \prior d\boldsymbol{\theta}.
\label{eq:marginalL}
\end{equation}
%
but we can rewrite this integral using the importance pdf $q(\params)$ as
\begin{subequations}
\begin{align}
\marginal & = \int_{\boldsymbol{\Theta}} \lh \prior d\boldsymbol{\theta} \\
                & = \int_{\boldsymbol{\Theta}} \frac{\lh \prior}{q(\params)} q(\params) d\boldsymbol{\theta}, \\
                & = \int_{\boldsymbol{\Theta}} \frac{\target}{q(\params)} q(\params) d\boldsymbol{\theta}, \\
                & = \int_{\boldsymbol{\Theta}} w(\params) q(\params) d\boldsymbol{\theta}. 
\end{align}
\label{eq:marginalLIS}
\end{subequations}
Then, using a standard Monte Carlo approximation,
\begin{equation}
\marginal \approx { \widehat{Z}}= \frac{1}{N} \sum\limits_{i = 1}^{N} w(\params_i), \qquad \params_i\sim q(\params), \quad i=1, \ldots, N.
\label{eq:marginalMC}
\end{equation}
This is the arithmetic mean of the {unnormalised} importance weights. It is important to remark that Eq.~\eqref{eq:marginalMC} is only valid if $q(\params)$ is normalised. 
}


\subsection{Summary of the importance sampling algorithm}

The method is summarised in Algorithm \ref{al:IS} \citep{Rubin1987}. 

{The set of weighted samples $\{ \boldsymbol{\theta}_i, \bar w_i \}_{i=1}^N$ enables us to approximate posterior expectations of the form 
$$
I(g) := \mbE[ g(\boldsymbol{\theta}) | \mathbf{y} ]
$$ 
for any integrable function $g(\boldsymbol{\theta})$. In particular, it is straightforward to construct the weighted-mean estimator 
\begin{equation}
\widetilde{I}_N(g) := \sum_{i=1}^N g(\boldsymbol{\theta}_i) \bar w_i
\label{eqI(N)}
\end{equation} 
and it can be proved, under mild assumptions, that $\lim_{N\to \infty} \widetilde{I}_N(g)=I(g)$ (see, e.g., \citet{Robert2005}).
}


\begin{algorithm}
{
\SetAlgoNoLine 
\tcc{{\bf Initialization:}}
- Choose the number of samples, $N$.\\
- {Choose a proposal density, $q(\params)$}.\\ 
\tcc{{\bf Sampling:}}
- Draw $\params_{1},...,\params_{N} \sim   {q(\params)}$\;
       - Compute the   {unnormalised} weights 
       $$
       w(\params_{i})=\frac{\pi(\params_{i}|{\bf y} )}{{q(\params_i)}}.
       $$ 
       for $i=1,..,N$\;
       \tcc{{\bf Output:}}
  {- Return weighted samples $\{\params_{i},w(\params_{i})\}_{i=1}^N$ in order to construct estimators of the $\widetilde I_N(g)$ in Eq. \eqref{eqI(N)}. More specifically, one can also obtain the estimators $\widetilde{I}_N$ in Eq. \eqref{eq:NormBayesianMonteCarloIS} and $\widehat{Z}$ in Eq. \eqref{eq:marginalMC}.}
 \caption{Generic Importance Sampling}
 \label{al:IS}
 }
\end{algorithm}

\section{{Adaptive Target} Adaptive Importance Sampling (ATAIS)}
\label{sec:ATAIS}

\subsection{Adaptive Importance Sampling}
\label{sec:AIS}

The generic algorithm described in Section~\ref{sec:BayesianInference} works fine when the dimension of $\params$ is not high. If the number of parameters to infer is large enough, IS may fail in estimating the expected value of $\params$ or obtaining  {its} approximated posterior distribution. The reason is  {that those} samples drawn from the proposal distribution may not cover the parameter space correctly. In high-dimensional problems,  {a large number of samples might be required to get accurate results}. A solution to this issue is to repeat the process after adapting the proposal using the information given by the importance weights \citep[see][for an extensive review]{Bugallo2017}. This iterative process is called Adaptive Importance Sampling \citep[AIS;][]{Oh1992}. 

There exist many procedures for adapting the proposal in AIS \citep{Cappe2004,Cornuet2012,Elvira2015,Martino2017}. The standard scheme consists in updating  {the mean and variance} of the proposal distribution  {with} the mean and variance from Eqs.~\ref{eq:NormBayesianMonteCarloIS} and \ref{eq:VarBayesianMonteCarloIS}. A classic alternative is to  {replace} the mean by the maximum a posteriori $\params_\mathrm{MAP}$. This $\params_\mathrm{MAP}$ is the set of parameters that maximises the target distribution $\target$. For symmetric posterior distributions like multivariate gaussians, $\params_\mathrm{MAP}$ is equal to the expected value $\mbE \left[ \params | \mathbf{y} \right]$. However,  {this does not hold for asymmetric or multimodal distributions.} The choice of $\params_\mathrm{MAP}$ or $\mbE\left[ \params | \mathbf{y} \right]$ depends on the {problem}.  {When the target distribution is multimodal, for instance, the use of $\params_\mathrm{MAP}$ leads to more samples tracing the highest mode, which might leave the remaining ones poorly identified or even missing.} Instead, the use of $\mbE \left[ \params | \mathbf{y} \right]$ will probably sample all the modes but it may need many iterations to obtain a good representation of each single mode.  {Algorithm~\ref{al:AIS} provides pseudocode for a general implementation of AIS.}

\begin{algorithm}
{
\SetAlgoLined 
\tcc{{\bf Initialization:}}
- Choose the number of iterations, $T$\;
- Choose the number of samples per iterations, $N$\;
- {Choose an initial proposal $q(\params|\boldsymbol{\beta}_1)$, where $\boldsymbol{\beta}_1$ is the set of proposal parameters. For example, if $q(\params|\boldsymbol{\beta}_1)$ is a multivariate Gaussian pdf, then $\boldsymbol{\beta}_1 = \{ \boldsymbol{\mu}_1, \boldsymbol{\Sigma}_1 \}$, where $\boldsymbol{\mu}_1$ is the mean vector and $\boldsymbol{\Sigma}_1$ is the covariance matrix of the Gaussian distribution.} \\
 \tcc{{\bf Iterations:}}
 \For{t = 1, ..., T}{
  \tcc{{\bf Sampling:}}
       - {Draw $\params_{1,t}, \ldots, \params_{N,t} \sim q(\params|\boldsymbol{\beta}_t)$}\;
       - Compute the unnormalised weights 
       $$
       w(\params_{i,t})=\frac{\pi(\params_{i,t}|{\bf y} )}{{q(\params_{i,t}})}.
       $$ 
       for $i=1,..,N$\;
       \tcc{{\bf Adaptation of proposal:}}
         - {Compute a new parameter set $\boldsymbol{\beta}_{t+1}$ for the proposal distribution, using the information of the weighted samples}\;
  }
  \tcc{{\bf Output:}}
  - Return the $NT$ weighted samples. $\{\params_{i,t},w(\params_{i,t})\}_{i=1}^N$ with $t=1,..,T$.
 \caption{Generic Adaptive Importance Sampling}
 \label{al:AIS}
 }
\end{algorithm}

{
The final estimators of a generic AIS algorithm can be expressed as
{
\begin{equation}
p({\bf y}) \approx \widehat{Z}_{T,N} = \frac{1}{NT}\sum_{t = 1}^{T} \sum_{i = 1}^{N}  w(\params_{i,t}),
\label{Z_AIS}
\end{equation}
\begin{equation}
\mbE \left[ \params | \mathbf{y} \right] \approx \widetilde{I}_{T,N} = \frac{1}{NT \widehat{Z}}\sum_{t = 1}^{T} \sum_{i = 1}^{N} \params_{i,t} w(\params_{i,t}).
\label{eq:NormBayesianMonteCarlo_AIS}
\end{equation}
}
%
}

\subsection{Weight clipping}
\label{sec:clipping}

 {Algorithm~\ref{al:AIS} {can be enhanced using different techniques}. A convenient modification of AIS} that is useful for high-dimensional problems is the so-called \emph{weight clipping} {\citep{Koblents2014,Miguez2018}}. A usual problem when working with AIS is the degeneracy of the importance weights due to the curse of dimensionality \citep[see][]{Bengtsson2008}. If the number of parameters to be inferred and/or the space to be explored are large, it is difficult to reach {regions of the state space with high probability density} during the first iterations. As a result, all or most of the samples drawn from the proposal will have null weight. Therefore, the adaptation will be done with only one sample or very few samples, which makes the method inefficient. A solution to this problem is to select a fixed number of samples with the largest weights and assign {all of them the same weight.} The expected value and the variance of the samples are determined using the new weights and the adaptation is modified consequently. {The resulting scheme preserves the convergence properties of classical IS while mitigating the weight degeneracy problem.}


{
\subsection{AIS with adaptive target}
\label{sec:AT}
In Bayesian  {inference}, it is  {common} to deal with noisy data. The lack of precise information about the {power} of this noise conditions the results of the inference method, since an incorrect choice of {this parameter} {may hamper the convergence of the algorithm}. Inferring both the set of parameters entering the {map $\mathbf{f}(\cdot)$} and the noise {power ${\sigma}^2$} is difficult. We propose a method to {optimise the value of $\sigma^2$ while inferring the parameters using AIS}. 

As mentioned in Section~\ref{sec:bayesFit}, the {map} $\mathbf{f}(\cdot)$ is a function of the parameters $\boldsymbol{\lambda}$, such that
\begin{eqnarray}
{\bf f}(\boldsymbol{\lambda})=[f_1({\bm \lambda}),...,f_{d_\lambda}({\bm \lambda})]^{\top}: {\bm \Lambda} \subseteq \mathbb{R}^{d_\lambda} \rightarrow   \mathbb{R}^{d_y}, 
\end{eqnarray}
where $\boldsymbol{\Lambda}$ is the $\boldsymbol{\lambda}$-parameter space. If the perturbation noise is assumed to follow a normal distribution,
\begin{eqnarray}
\mathbf{e}=[e_1,...,e_{d_\lambda}]^{\top} \sim \mathcal{N}({\bf e}|{\bf 0}, \sigma^2 {\bf I}_{d_\lambda}),
\end{eqnarray}
{where 
$\sigma^2$ is the noise power. Then, the likelihood  function is}
%
\begin{eqnarray}
 p({\bf y}|\boldsymbol{\lambda},\sigma^2)&=&\frac{1}{(2\pi\sigma^2)^{d_\lambda/2}}\exp\left( -\frac{1}{2\sigma^2} ||\mathbf{y} - {\bf f}(\boldsymbol{\lambda}) ||^2\right) 
\label{eq:LH}
\end{eqnarray}
Note that we have two types of variables of interest: the vector $\boldsymbol{\lambda}$ contains the parameters of the nonlinear {map} ${\bf f}(\boldsymbol{\lambda})$, whereas $\sigma$ is a scale parameter of the likelihood function. In our approach, AIS is used to infer the values of $\boldsymbol{\lambda}$, while {$\sigma^2$} is adapted at each iteration according to the quadratic difference of the best fit. The latter is given by the approximate MAP estimator, $\boldsymbol{\lambda}_\mathrm{MAP}$ (see Algorithm~\ref{al:ATAIS}).
\begin{equation}
\widehat{\sigma}_{\mathrm{MAP}}^2 =\frac{1}{{d_y}}  \sum_{k=1}^{{d_y}} (y_k-f_k(\boldsymbol{\lambda}_\mathrm{MAP}))^2, 
\label{eq:sigmaMAP}
\end{equation}
}
Note that {$\widehat{\sigma}_\mathrm{MAP}^2$} needs to be initialised at step $i = 1$ and it is updated only if {$\widehat{\sigma}_i^2 < \widehat{\sigma}_{i-1}^2$} (where $\widehat{\sigma}_i^2$ is the variance at the step $i$). Therefore, {$\widehat{\sigma}^2$} corresponds to a minimum square error (MSE). An outline of the method {with multivariate Gaussian proposals} is given in  {Algorithm~\ref{al:ATAIS}}. {Empirical results show that it is always convenient to start with a large value of {$\widehat{\sigma}^2$}}. 

\begin{algorithm}
{
\SetAlgoLined 
\tcc{{\bf Initialization:}}
Choose $N$, ${\bm \mu}_1$, ${\bm \Sigma}_1$, and {initialise $\widehat{\sigma}_{\texttt{MAP}}>>1$ and $\pi_{\texttt{MAP}} = 0$} \;
 \tcc{{\bf Iterations:}}

 \For{t = 1, ..., T}{

  \tcc{{\bf Sampling:}}
   - Draw $\Lparams_{1,t},...,\Lparams_{N,t} \sim q(\Lparams)$\; 
   - Weight the samples according to\ 
      \begin{eqnarray*}
        w_{n,t}=\frac{\pi(\Lparams_{n,t}|\widehat{\sigma}_{\texttt{MAP}},{\bf y})}{q(\Lparams_{n,t})},  
      \end{eqnarray*}
      with $n=1,...,N$\;
  \tcc{{\bf Current MAP estimation:}}
   - Obtain $\widehat{\Lparams}_t = \arg\max\limits_{1 \le n \le N} \pi_t(\Lparams_{n,t})$,  and compute $\widehat{\bf{r}}_t={\bf f}(\widehat{\Lparams}_t)$\; 
   - Compute $\widehat{\sigma}_t^2 = {\frac{1}{d_{\Lparams}}||\mathbf{y} - \mathbf{f}(\widehat{\Lparams}_t)||^2}$\;

 \tcc{{\bf Global MAP estimation:}}
   - If  $\widehat{\sigma}_t^2 \leq \widehat{\sigma}_{\texttt{MAP}}^2$, then set $ \widehat{\sigma}_{\texttt{MAP}}^2 =\widehat{\sigma}_t^2$\;
   - If  $\pi_t(\widehat{\Lparams}_t) \geq\pi_{\texttt{MAP}}$, then set $\widehat{\Lparams}_{\texttt{MAP}}=\widehat{\Lparams}_t$ and   $\pi_{\texttt{MAP}}=\pi_t(\widehat{\Lparams}_t)$\;
    
 \tcc{{\bf Adaptation:}}
   - Set\
   \begin{eqnarray*}
    {\bm \mu}_t&=&\widehat{\Lparams}_{\texttt{MAP}}, \\
    {\bm \Sigma}_t&=&\sum_{n=1}^N \bar{w}_{n,t} (\Lparams_{n,t} - \bar{\Lparams}_{\texttt{MAP}})^{\top} (\Lparams_{n,t}-\bar{\boldsymbol{\lambda}}_{\texttt{MAP}}) + \beta {\bf I}_{d_y},
   \end{eqnarray*}
  where  ${\bar w}_{n,t}=\frac{w_{n,t}}{\sum_{i=1}^N w_{i,t}}$ are the normalised weights, $\bar{\Lparams}_t=\sum_{n=1}^N \bar{w}_{n,t} \Lparams_{n,t}$ and $\beta >0$\;
  }
  \tcc{{\bf Output:}}
  - Return the MAP estimators, and all the weighted samples $\{\Lparams_{n,t},\widetilde{w}_{n,t}\}_{n=1}^N$ for all $t$, with the corrected weights\
  \begin{equation*}
   \widetilde{w}_{n,t}=w_{n,t} \frac{\pi_{T+1}(\Lparams_{n,t})}{\pi_t(\Lparams_{n,t})}
  \end{equation*}
 \caption{ATAIS: AIS with adaptation of the target}
 \label{al:ATAIS}
 }
\end{algorithm}

\section{Model selection with ATAIS}
\label{sec:ModelSelection}

{We include here an example that demonstrates the capability of ATAIS in accurately computing the marginal likelihood within a model selection problem. ATAIS results are compared with those obtained by numerically integrating the marginal likelihood using a dense grid. In addition, we compare ATAIS with nested sampling methods. For this purpose, we have chosen UltraNest \citep{Buchner2021b}, an implementation of a rigorous nested sampling method. For the sake of reproducibility, we have used the example described in the UltraNest tutorial\footnote{The code can be downloaded from the UltraNest GitHub page (https://johannesbuchner.github.io/UltraNest/example-sine-mo\-del\-com\-pa\-ri\-son.html).}. The problem consists in computing the marginal likelihood for a given dataset using two distinct models. {The first} model ($M_0$) is, 
}
\begin{equation}
\mathbf{y} = B + \boldsymbol{\epsilon}.
\end{equation}
The {second} model ($M_1$) is a one dimensional sinusoid,
\begin{equation}
\mathbf{y} = A_1 \sin \left( 2\pi \left(\frac{\mathbf{t}}{P_1} + t_1 \right)\right) + B + \boldsymbol{\epsilon}.
\end{equation}
{In both models} $\boldsymbol{\epsilon} \sim \mathcal{N}(0,\sigma)$ represents white, gaussian noise. 
The dataset consists of 50 points randomly selected between $t = 0$ and $t = 5$ from model $M_1$ (meaning M1 is the \emph{true} model). The model parameters are fixed to $B=1$ for the baseline of the signal, $A_1=0.9$ for the amplitude, $P_1 = 3$ for the period and $t_1=0$ for the phase. {The variance of the noise is set to $\sigma^2_\epsilon = 1$.} 

\subsection{Comparison with numerical integration of the marginal likelihood}

{For the numerical integration, the parameter space has been divided into cells of {uniform} width 0.1 for $A_1$, $P_1$ and $B$, and 0.02 for $t_1$. The limits of the integral are the same as the boundaries of the prior pdfs defined in the UltraNest tutorial example: $B \in [-10, 10]$, $A_1 \in [0.1, 100]$, $P_1 \in [0.3, 30]$, and $t_1 \in [0, 1]$. We have applied an expensive trapezoidal rule for the integration in order to obtain the ground-truth values. The marginal likelihood computed\footnote{The values of $\log Z$ listed in the text have been obtained by using the likelihood function defined in the example of the UltraNest tutorial ($\chi^2$). For a normalised Gaussian likelihood function, we obtain $\log Z_{M_1} = -82.28$ and $\log Z_{M_0} = -81.69$.} for the model $M_1$ is $\log Z_{M_1} = -36.33$, while we obtain $\log Z_{M_0} = -35.74$ for  $M_0$. With these values, the Bayes factor $K = e^{-36.33 + 35.74} = 0.55$. Therefore, {the preferred model with the uniform priors defined in this example} is $M_0$, despite the data {having} been generated with the model $M_1$. Actually, this result is not surprising. The signal-to-noise ratio of the data is low since the amplitude of the sinusoid ($A_1$) used to generate the data is lower than the standard deviation of the noise. Note that the marginal likelihood tends to penalise models with high complexity.}

{We have applied ATAIS to {the simulated dataset}. We have iterated the algorithm 20 times, each one with $10^4$ samples, for a total of $2 \times 10^5$ samples in the single run. {The prior pdfs defined here are uniform, with the same ranges in the parameter space than those used for the numerical integration}. The proposal pdf of each model parameter is assumed normal with initial variance $\sigma^2 = 1$. The initial mean of each proposal pdf has been chosen randomly from the corresponding prior pdf. The run has been completed in 5.8 seconds in a 2.3 GHz Quad-Core Intel Core i5 processor. With this configuration for ATAIS, we have obtained $\log Z_{M_1} = -36.42$, $\log Z_{M_0} = -35.74$ and $K = 0.51$. Therefore, ATAIS approximates correctly the true values of the marginal likelihood (see above the results of expensive trapezoidal integration). No substantial difference is found in these values with distinct runs of ATAIS.} 

{ATAIS performs inference over the model parameters and the noise variance ($\widehat{\sigma}_{\epsilon}^2$). Eventually, a marginal likelihood can be computed for different values of $\sigma_\epsilon$ ($Z(\sigma_\epsilon)$). The later is done by sampling from a prior $p(\sigma_\epsilon)$ and applying Eqs.~24 and 25 in \citet{Martino2021}. Figure~\ref{fig:cornerPlot} shows a corner plot of the marginal posteriors and pairwise correlations for the model $M_1$. The parameters inferred by ATAIS are shown at the top of each histogram, together with the 90\% credibility interval. In this example, they correspond to the MAP estimators. {Hereafter, we use $\widehat{\sigma}_{MAP}$ instead of $\widehat{\sigma}_{\epsilon}^2$}. The algorithm converges to $\widehat{\sigma}_{\mathrm{MAP}}^2 = 0.822$ for this model. The variation of the marginal likelihood with the value of the variance for the model $M_1$ (${\sigma_{M_1}}$) is shown in Figure~\ref{fig:MarginalZ}. Similarly, Figure~\ref{fig:MarginalB} shows the marginal posterior pdf of the constant $B$ in the model $M_0$. For this model, the  algorithm converges to $\widehat{\sigma}_{\mathrm{MAP}}^2 = 1.27$. The variation of the marginal likelihood with the variance for this model (${\sigma_{M_0}}$) is also shown in Figure~\ref{fig:MarginalZ}.}

\begin{figure}
   \centering
   \includegraphics[width=\columnwidth]{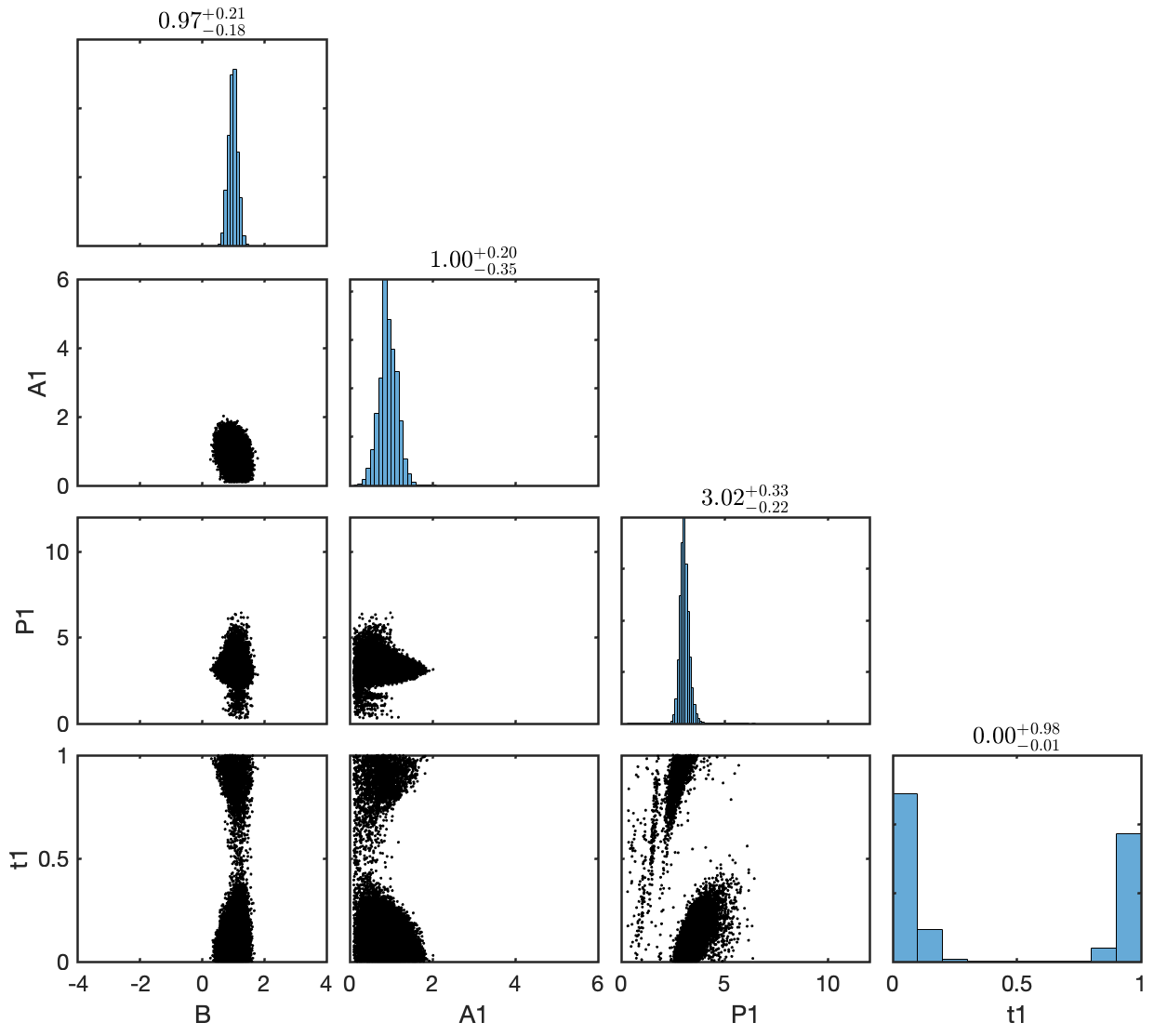} 
   \caption{Corner plot of the parameters and the marginal posterior pdfs of model $M_1$ obtained with 
   ATAIS. The bin in each histogram is 0.1.}
   \label{fig:cornerPlot}
\end{figure}

\begin{figure}
   \centering
   \includegraphics[width=\columnwidth]{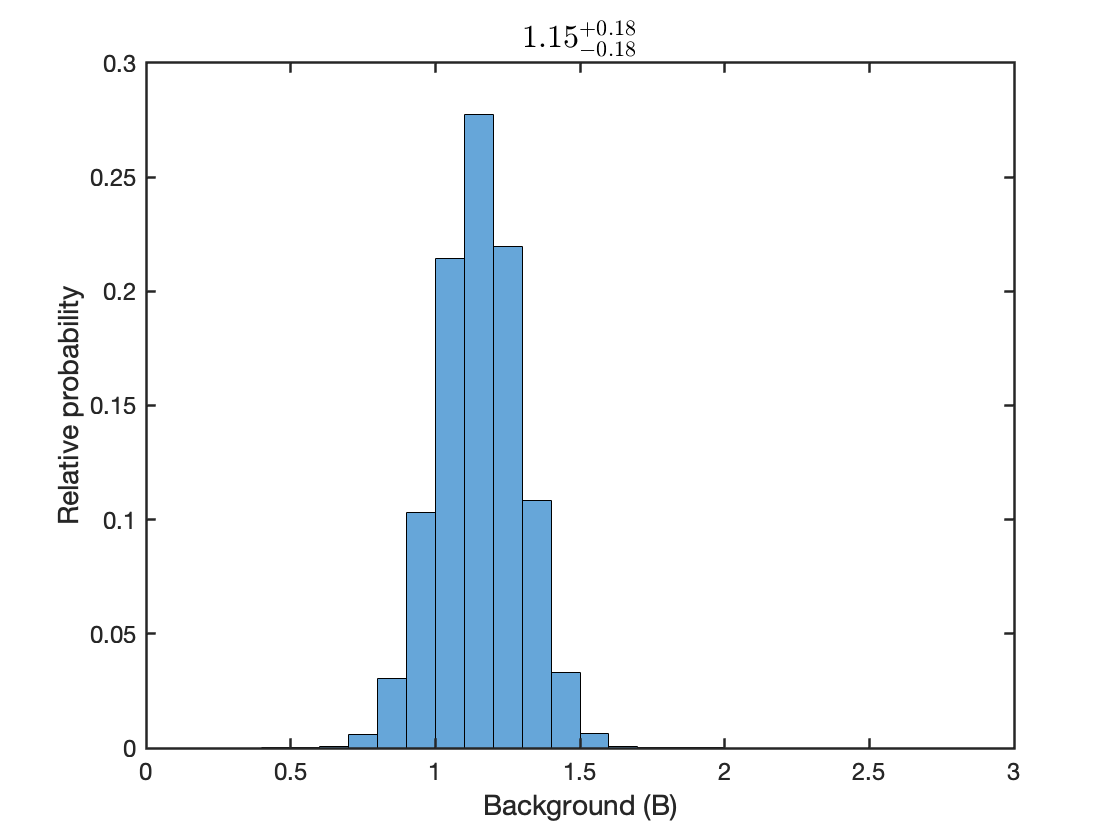} 
   \caption{Marginal posterior pdf of the constant $B$ in model $M_0$ obtained with 
   ATAIS. The bin in each histogram is 0.1. The inferred MAP and the 90\% credibility interval
   are shown at the top.}
   \label{fig:MarginalB}
\end{figure}

\begin{figure}
   \centering
   \includegraphics[width=0.48\columnwidth]{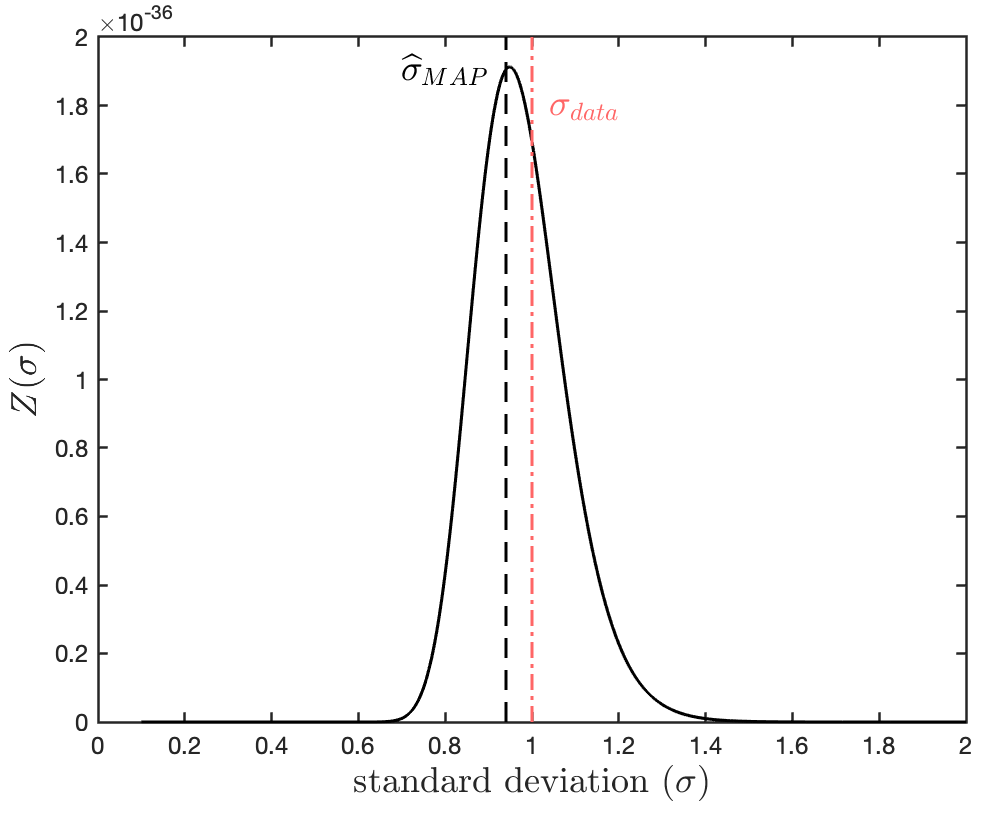} 
   \includegraphics[width=0.48\columnwidth]{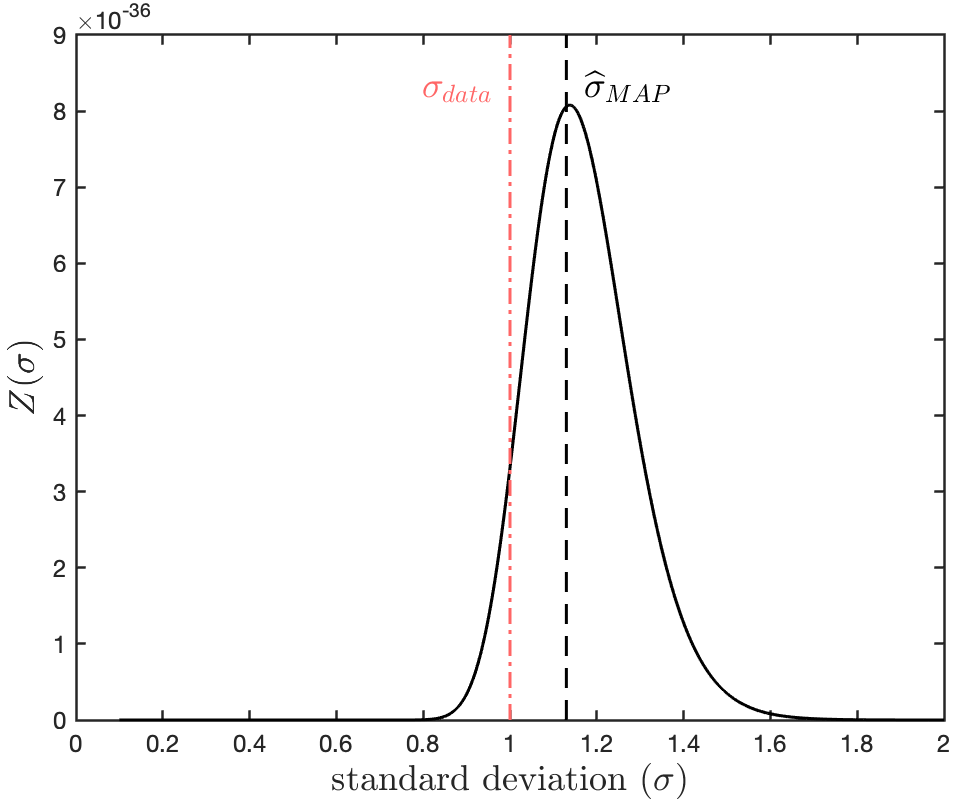} 
   \caption{Marginal likelihood for each value of ${\sigma}$ for the model $M_1$ 
   (left) and the model $M_0$ (right). The dashed line marks the value of $\widehat{\sigma}_{\mathrm{MAP}}$
   determined by the algorithm for the corresponding model and the dotted-dashed line is the value  
   used for the simulation ($\sigma = 1$).}
   \label{fig:MarginalZ}
\end{figure}

\subsection{Comparison with nested sampling}

 {The marginal likelihoods computed by UltraNest for the dataset and models used in the previous section are $\log Z_{M_1} = -32.99$ and $\log Z_{M_0} = -35.64$. Thus, the Bayes factor is $K = 14.17$.\footnote{There is a slight mistake in the UltraNest tutorial. The prior for the parameter $P1$ is defined between 1 and 100, instead of 0.3 and 30 as indicated in the comment line. We have corrected it for our test.}} {With this result, the preferred model is $M_1$. Note that the prior pdfs defined for the parameters $A_1$ and $P_1$ in the tutorial are uniform in a logarithmic scale. As a consequence, the estimation of the marginal likelihood is different from that computed using the numerical integration with a uniform grid.} The algorithm yields this result when 400 live points (or nodes) are used. The total number of evaluations of the likelihood for the model $M_1$ is $\sim 1.8 \times 10^5$.} 

{For the comparison with ATAIS, we have defined the same prior pdfs with the model $M_1$. Like in the previous section, the proposal pdf of each model parameter is assumed normal with initial variance $\sigma^2 = 1$. {We have obtained $\log Z_{M_1} = -31.14$}, while the marginal likelihood of $M_0$ remains the same than in the previous section ($\log Z_{M_0} = -35.74$). Therefore, the Bayes factor is $K \approx 99$. Like with UltraNest, the preferred model with these prior pdfs is $M_1$. The results with ATAIS are in agreement with those obtained using UltraNest.} {The difference between UltraNest and ATAIS results is likely due to the way that ATAIS deals with the noise variance, which in UltraNest is fixed.}

{The inferred values for the four parameters of $M_1$ using the log-uniform priors for $A_1$ and $P_1$ are similar to those obtained in the previous section. Their corresponding marginal posteriors are shown in the diagonal of Figure~\ref{fig:cornerPlot2}. The values on top of those marginal posteriors are the estimates of the parameters. The lower and upper limits are the 90\% credibility intervals. The algorithm converges to $\widehat{\sigma}_{\mathrm{MAP}}^2 = 0.822$ for this model. This value is similar to that obtained in the previous section. Our results show that the ATAIS scheme is robust and that the inference is not dependent of the prior pdfs provided the correct values are not excluded.}

\begin{figure}
   \centering
   \includegraphics[width=\columnwidth]{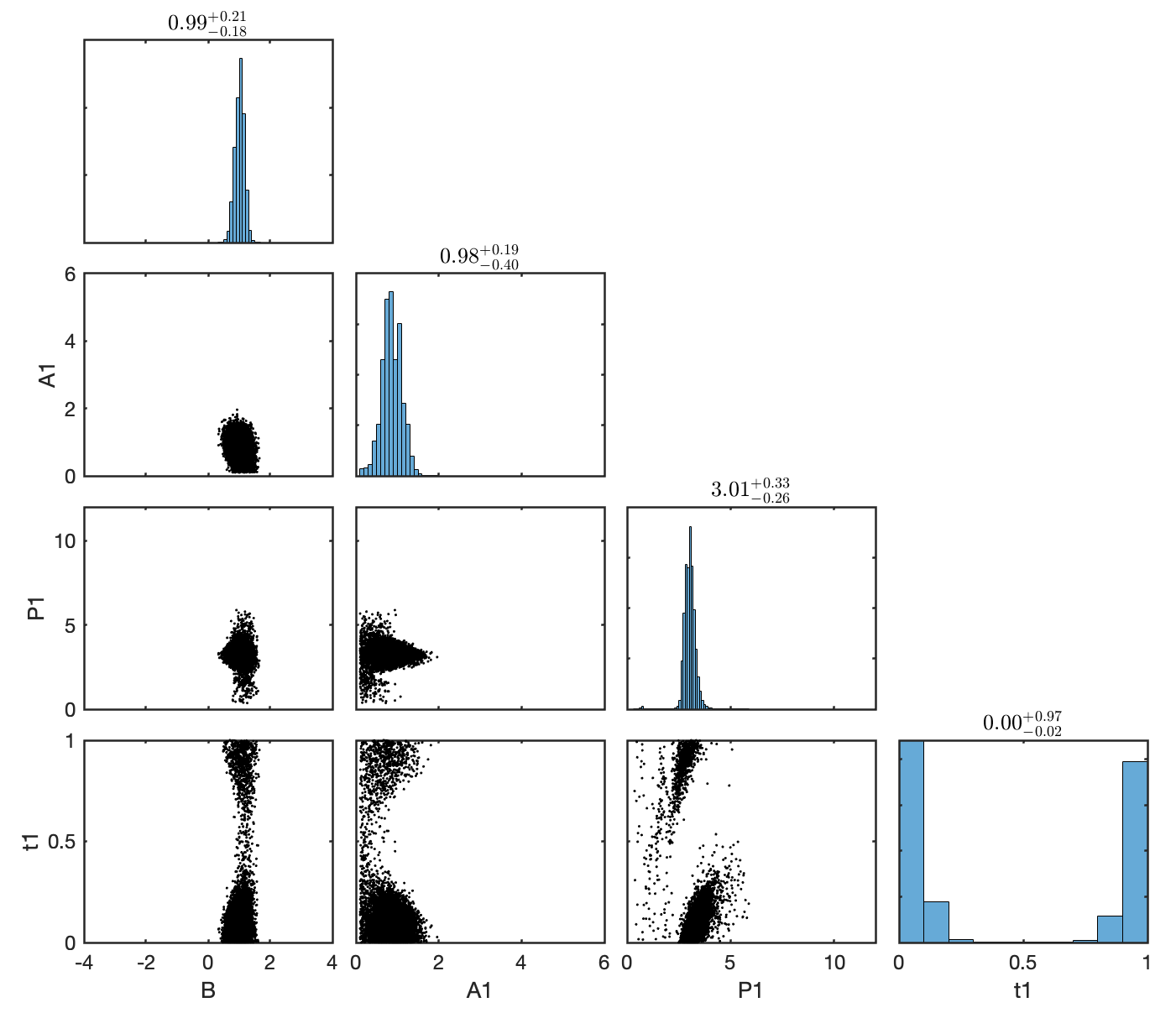} 
   \caption{Corner plot of the parameters and the marginal posterior pdfs of model $M_1$ obtained with 
   ATAIS with log-uniform prior pdfs for $A_1$ and $P_1$. The bin in each histogram is 0.1.}
   \label{fig:cornerPlot2}
\end{figure}

\section{An application of ATAIS for bayesian inference: flare light curves}
\label{sec:inference}

\subsection{Simulated data of a flare with oscillation}
\label{sec:Simul}

{The performance of our method for Bayesian inference and model selection is tested against simulated data of a flare light curve with a damped oscillation. The use of simulated data permits to control the sources of error in a way that it is not possible with real data. We assume the flare emission is governed by a short exponential rise phase followed by a longer exponential decay. The rise phase is modelled by 
\begin{equation}
\label{eq:rise}
\mathbf{y}_r = C e^{(\mathbf{t} - t_p)/\tau_r},~~\forall \mathbf{t} \le t_p.
\end{equation}
Here, $\mathbf{t}$ is a vector of time instants, $C$ is a constant related to the flare amplitude, $\tau_r$ is the rise time and $t_p$ is the time of the peak emission. The decay phase is represented by 
\begin{equation}
\label{eq:decay}
\mathbf{y}_d = C e^{-(\mathbf{t} - t_p)/\tau_d},~~\forall \mathbf{t} > t_p,
\end{equation}
where $\tau_d$ is the decay time. The oscillation is modelled by the sinusoid
\begin{equation}
\label{eq:osc}
\mathbf{y}_o = A \sin \frac{2\pi}{P} ( \mathbf{t} - t_i ),~~\forall \mathbf{t} \ge t_i,
\end{equation}
where $A$ is the amplitude of the oscillation, $P$ is its period and $t_i$ its starting time. Finally, the oscillation is exponentially damped by
\begin{equation}
\label{eq:damping}
\mathbf{y}_e = e^{-(\mathbf{t} - t_i)/\tau_e},~~\forall \mathbf{t} \ge t_i,
\end{equation}
The combination of Eqs.~\ref{eq:rise} to \ref{eq:damping} is a generic model for the flare light curve with an oscillation,
\begin{equation}
\label{eq:flareModel}
\mathbf{y} = \mathbf{y}_r + \mathbf{y}_d + \mathbf{y}_o \, \mathbf{y}_e.
\end{equation}
Therefore, the model contains eight parameters to be inferred,
\begin{equation}
\label{eq:flareParams}
\boldsymbol{\lambda} = [C, t_p, \tau_r, \tau_d, A, P, t_i, \tau_e].
\end{equation}
For the simulations, we have used the values listed in Table~\ref{tab:simul}. For simplicity, we have considered gaussian white noise. The variance of this noise has been fixed to $\sigma^2 = 4$ (in flux units). Figure~\ref{fig:simul} shows the simulated light curve. Note that the algorithm does not require prior information about units in the model. Therefore, we do not include them in the figure or the table. 
%
\begin{table}
   \centering
   \caption{Values of the model parameters used in the simulation of the flare light curve. 
   }             
   \label{tab:simul}
   \begin{tabular}{cccccccc} 
   \hline
   $C$ & $t_p$ & $\tau_r$ & $\tau_d$ & $A$ & $P$ & $t_i$ & $\tau_e$ \\
   \hline
   72 & 30.2 & 5 & 30 & 20 & 9 & 24 & 40 \\
   \hline
   \end{tabular}
\end{table}
%
\begin{figure}
   \centering
   \includegraphics[width=\columnwidth]{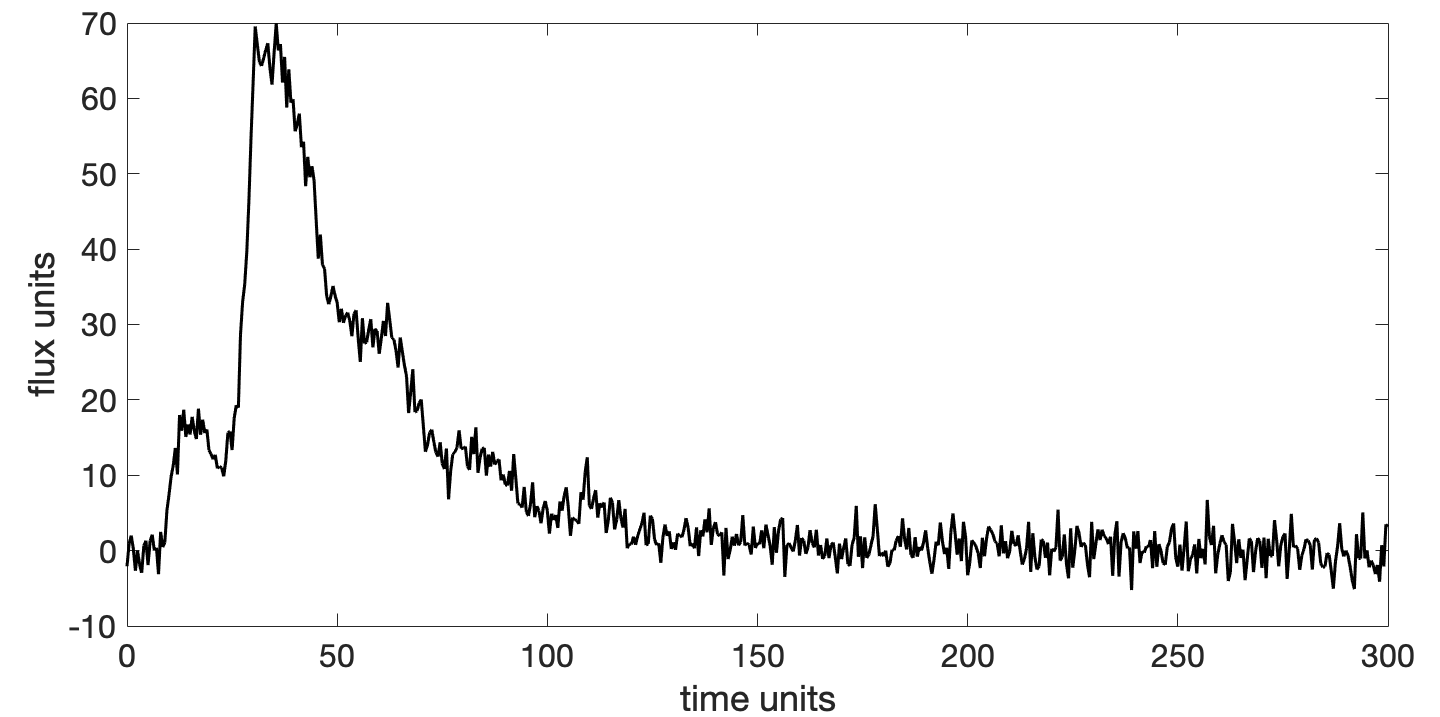} 
   \caption{Simulated flare light curve with damped oscillation used in this work.}
   \label{fig:simul}
\end{figure}
%
}

{We propose two alternatives for model testing. The first alternative is a model of several flares (each with an exponential rise and decay phases) that account for a process of repetitive ignition of different loops. The second alternative is the model of a single flaring loop with a damped oscillation described previously. {Our main goal is to perform inference on the parameters of the two models. Additionally, we} estimate the model evidence $\widehat{Z}$ (Eq.~\ref{eq:marginalMC}) for each model and compare them using ATAIS. To check whether the algorithm is able to distinguish between the correct model and a multi-loop model with many parameters, we have assumed up to three flaring loops in the multiple flares model. The model with two flaring loops contains eight parameters to infer, like the model of a single loop with an oscillation. The model with three flaring loops has 12 different parameters. Hereafter, we will denote by $M_1$ the model of the single flare with oscillation and $M_2$ and $M_3$ the models with two and three flares without oscillation, respectively. 
}

\begin{figure}
   \centering
   \includegraphics[width=\columnwidth]{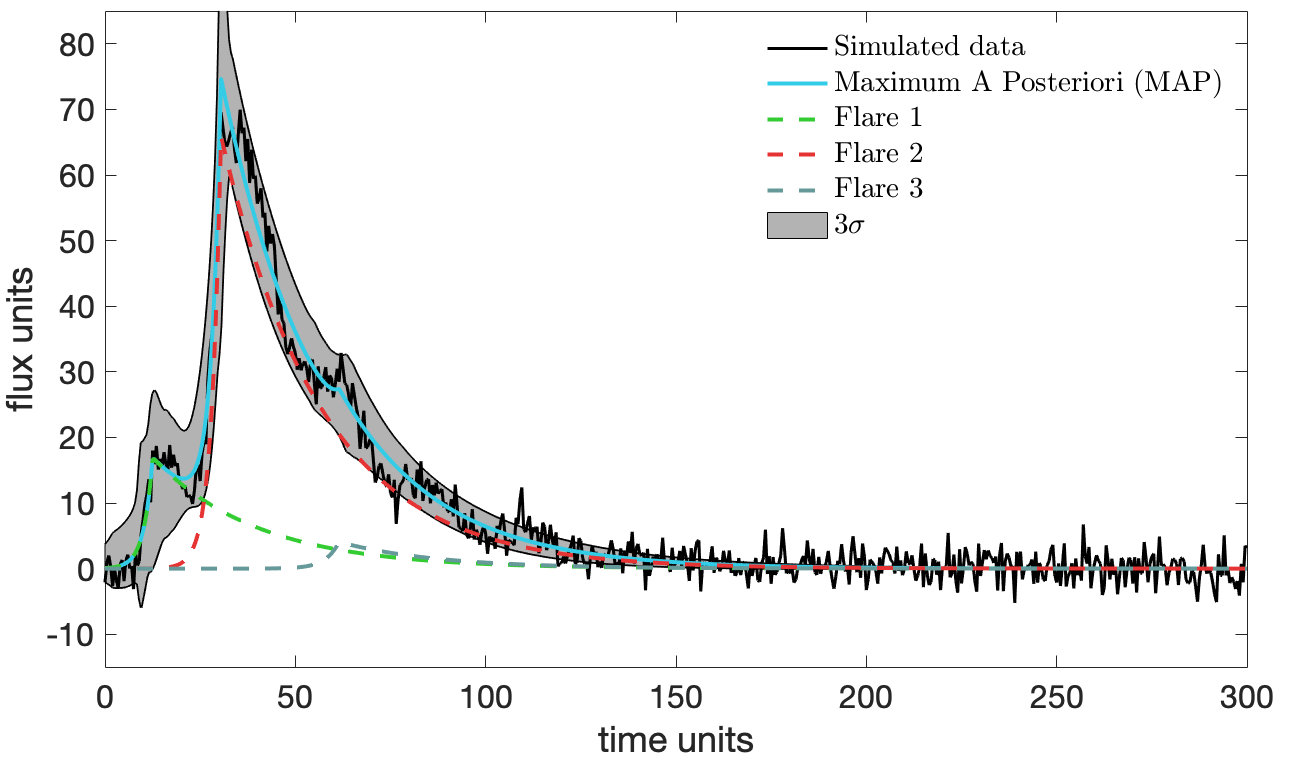} 
   \caption{Result of the inference over the model of a single loop with an oscillation. The curve generated 
   with the maximum a posteriori (MAP) estimate is plotted as a continuous (cyan) line. The dashed (red)
   line represents the exponential rise and decay phases and the dotted (green) line is the damped 
   oscillation. The filled area is the $3\sigma$ envelope. }
   \label{fig:Model1}
\end{figure}

\begin{figure}
   \centering
   \includegraphics[width=\columnwidth]{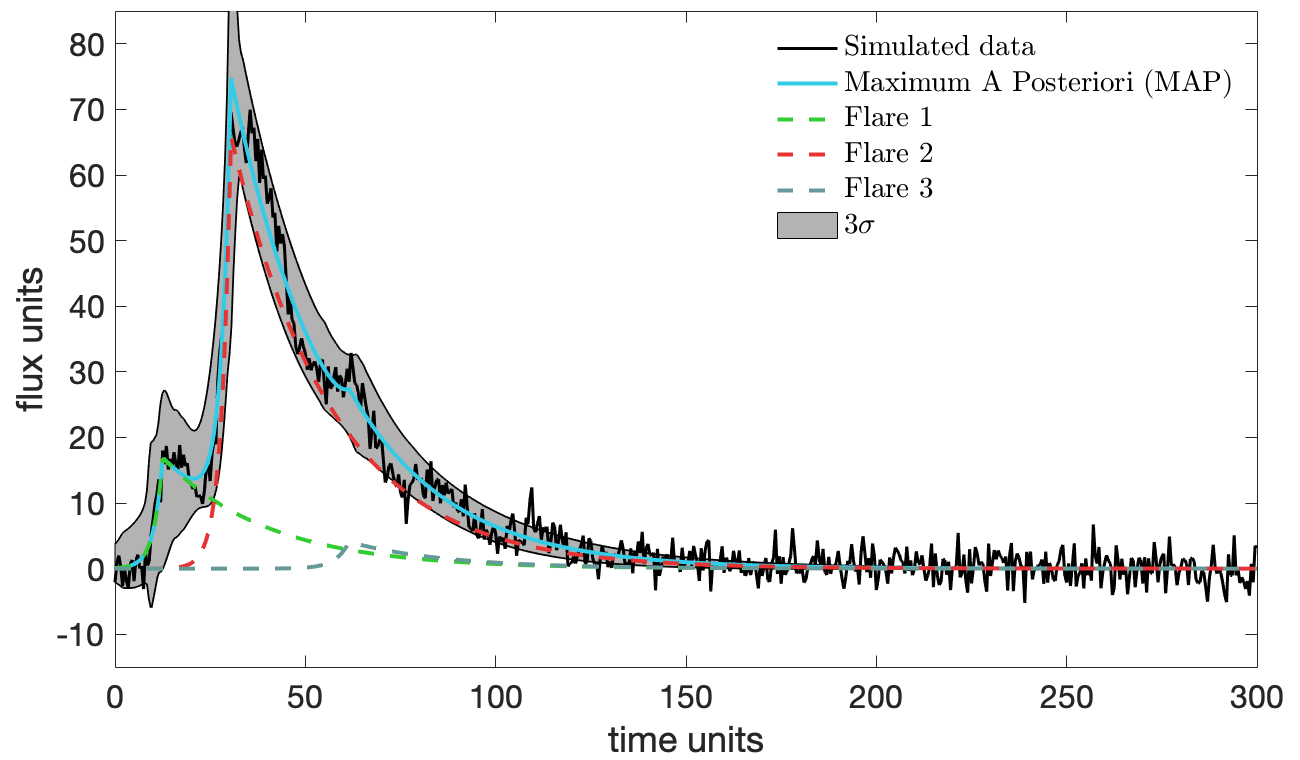} 
   \caption{Result of the inference over the model of three flaring loops. The curve generated 
   with the maximum a posteriori (MAP) estimate is plotted as a continuous (cyan) line. The dashed 
   lines represent the exponential rise and decay phases of each flaring event. The filled area is 
   the $3\sigma$ envelope. }
   \label{fig:Model3}
\end{figure}

Our algorithm has been run for each model using $10^4$ samples and $30$ iterations, for a total of $3 \times 10^5$ generated samples. {The prior pdfs for each model are shown in Table~\ref{tab:priors}.} The results of the inference are summarised in Figures~\ref{fig:Model1} and \ref{fig:Model3}. The former corresponds to the model of a single flaring loop with a damped oscillation ($M_1$). The latter shows the results for the multi-flare model with three flaring loops ($M_3$). The multi-flare model with two loops does not reproduce correctly any of the bumps in the simulated data beyond $t = 50$ and it is not considered here. In both figures, the continuous (cyan) line is the light curve generated with the maximum a posteriori (MAP) estimate and the grey area is the {$3\sigma$ envelope generated with the samples of every iteration of the algorithm}. At first glance, the two models seem to fit correctly the data. However, the estimations of the marginal likelihood of the two models give preference to the model of a single flare and damped oscillation ($\widehat{Z}_{M_1} / \widehat{Z}_{M_3} = 181$). {It is important to note that the value of the marginal likelihood depends on the selection of the prior pdfs. Different pdfs will result in a different Bayes factor}. Another indication that the preferred model is $M_1$ is the value of $\widehat{\sigma}_{\mathrm{MAP}}^2$ obtained by ATAIS. While for $M_1$, we obtain $\widehat{\sigma}_{\mathrm{MAP}}^2 = 1.03$, for $M_3$ the algorithm converges to $\widehat{\sigma}_{\mathrm{MAP}}^2 = 1.57$. Here, $\widehat{\sigma}_{\mathrm{MAP}}^2$ has units of squared flux in this example. Table~\ref{tab:results} shows the value of the parameters inferred for model $M_1$ with the $90\%$ confidence interval. The coincidence with the values used for the simulation (Table~\ref{tab:simul}) is noticeable. 

\begin{table}
   \centering
   \caption{Prior pdfs defined for each model. In the models $M_2$ and $M_3$, the superscript 
   indicates the flare component. 
   }             
   \label{tab:priors}
   \begin{tabular}{cc} 
   \hline
   Parameter & Prior  \\
   \hline
   \multicolumn{2}{c}{Model $M_1$}\\
   \hline
   $C$ & $\mathcal{U}(0,100)$  \\
   $t_p$ & $\mathcal{U}(25,50)$  \\
   $\tau_r$ & $\mathcal{U}(0,100)$  \\
   $\tau_d$ & $\mathcal{U}(0,150)$  \\
   $A$ & $\mathcal{U}(0,30)$  \\
   $P$ & $\mathcal{U}(0,50)$  \\
   $t_i$ & $\mathcal{U}(0,70)$  \\
   $\tau_e$ & $\mathcal{U}(0,300)$  \\
   \hline
   \multicolumn{2}{c}{Model $M_2$}\\
   \hline
   $C^{(1)}$ & $\mathcal{U}(0,100)$  \\
   $t_p^{(1)}$ & $\mathcal{U}(0,25)$  \\
   $\tau_r^{(1)}$ & $\mathcal{U}(0,100)$  \\
   $\tau_d^{(1)}$ & $\mathcal{U}(0,150)$  \\
   $C^{(2)}$ & $\mathcal{U}(0,100)$  \\
   $t_p^{(2)}$ & $\mathcal{U}(25,50)$  \\
   $\tau_r^{(2)}$ & $\mathcal{U}(0,100)$  \\
   $\tau_d^{(2)}$ & $\mathcal{U}(0,150)$  \\   
   \hline
   \multicolumn{2}{c}{Model $M_3$}\\
   \hline
   $C^{(1)}$ & $\mathcal{U}(0,100)$  \\
   $t_p^{(1)}$ & $\mathcal{U}(0,25)$  \\
   $\tau_r^{(1)}$ & $\mathcal{U}(0,100)$  \\
   $\tau_d^{(1)}$ & $\mathcal{U}(0,150)$  \\
   $C^{(2)}$ & $\mathcal{U}(0,100)$  \\
   $t_p^{(2)}$ & $\mathcal{U}(25,50)$  \\
   $\tau_r^{(2)}$ & $\mathcal{U}(0,100)$  \\
   $\tau_d^{(2)}$ & $\mathcal{U}(0,150)$  \\  
   $C^{(3)}$ & $\mathcal{U}(0,100)$  \\
   $t_p^{(3)}$ & $\mathcal{U}(50,100)$  \\
   $\tau_r^{(3)}$ & $\mathcal{U}(0,100)$  \\
   $\tau_d^{(3)}$ & $\mathcal{U}(0,150)$  \\
   \hline
   \end{tabular}
\end{table}

\begin{table*}
   \centering
   \caption{Values of the inferred parameters for model $M_1$. The subscripts and superscripts indicate the 90\% confidence interval.  
   }             
   \label{tab:results}
   \begin{tabular}{cccccccc} 
   \hline
   $C$ & $t_p$ & $\tau_r$ & $\tau_d$ & $A$ & $P$ & $t_i$ & $\tau_e$ \\
   \hline
   72.66$_{-0.61}^{+0.43}$ & 30.30$_{-0.31}^{+0.07}$ & 4.93$_{-0.17}^{+0.10}$ & 29.16$_{-0.38}^{+0.78}$ & 26.24$_{-0.66}^{+0.63}$ & 8.94$_{-0.08}^{+0.30}$ & 24.07$_{-0.31}^{+0.04}$ & 29.17$_{-0.25}^{+2.52}$ \\
   \hline
   \end{tabular}
\end{table*}

\subsection{Real data: white light flare}
\label{sec:Kepler}

{To conclude with the tests, we include an example with real data from a white-light flare observed with \emph{Kepler} and analysed by \citet{Pascoe2020}. These authors use a Bayesian inference tool developed specifically for analysing flare light curves \citep[the Solar Bayesian Analysis Tool, SoBAT;][]{Anfinogentov2021}. SoBAT contains an MCMC algorithm for inference and an importance sampling algorithm to determine the marginal likelihood of the data and perform model selection. \citet{Pascoe2020} analyse several white light flares in their article. Here, we focus on the data of the star KIC~12156549. The results of \citet{Pascoe2020} are discussed in their Section~3.2. The authors do not include the prior pdfs for all the parameters in their models, nor they show the expectations for the inferred values. In addition, the implementation of each one of their models contains one dimension more than our implementation, which corresponds to the observed noise variance. As a result, we cannot compare our estimations directly with theirs. However, we can compare the estimation of the noise variance that the authors obtained for each model, which is included in their Figures~6 and 8. 
}

{For the comparison, we have implemented the multi-flare model with two and four flares, together with the type P oscillation model with spline envelope. All those models are described in detail in \citet{Pascoe2020} and we do not reproduce them here. The prior pdfs for the peak time of the distinct flares are given in Section~3.2 of that work and those of their amplitudes, rise and decay times are indicated in their Section~2. For the oscillatory signal, we use uniform prior pdfs defined by the intervals [0, 400], [200, 400] and [0, 400] for the amplitude, starting time and period, respectively. Note that the starting time interval is chosen such that the oscillation would be triggered by the second flare, as suggested by \citet{Pascoe2020}. For the spline envelope, we use three points with the initial time coincident with the oscillation starting time, and the interpolating and last point as free parameters. Their uniform priors are defined in the interval [0, 400] and [400, 600]. The value of the interpolating point has a uniform prior pdf in the interval [0, 10]. Summarising, we compare three different models with dimensions $8$, $16$ and $15$, respectively. 
}

{Our results are shown in Figure~\ref{fig:Kepler}. They have been obtained after 10 iterations with $10^5$ samples per iteration, for a total of $10^6$ samples. The multi-flare model with two flares does not reproduce the observed light curve correctly, like in \citet{Pascoe2020} and we do not show the result here. The noise standard deviation determined from ATAIS for the four flares model is $\widehat{\sigma}_{\mathrm{MAP}} = 0.028$. For the model with the oscillation, we obtain $\widehat{\sigma}_{\mathrm{MAP}} = 0.35$. These values are similar to those obtained by \citet{Pascoe2020}. With the prior pdfs previously defined, the estimations of the marginal likelihood of the two models give preference to the multi-flare model with four flares ($M_4$) against the model with two flares and an oscillation ($M_5$), $\log \widehat{Z}_{M_4} - \log \widehat{Z}_{M_5} = 14.6$. Although this result is similar to that of \citet{Pascoe2020}, the values of the marginal likelihood are not comparable because the authors do not give information for every prior pdf in their oscillation model. 
}

\begin{figure*}
   \centering
   \includegraphics[width=\columnwidth]{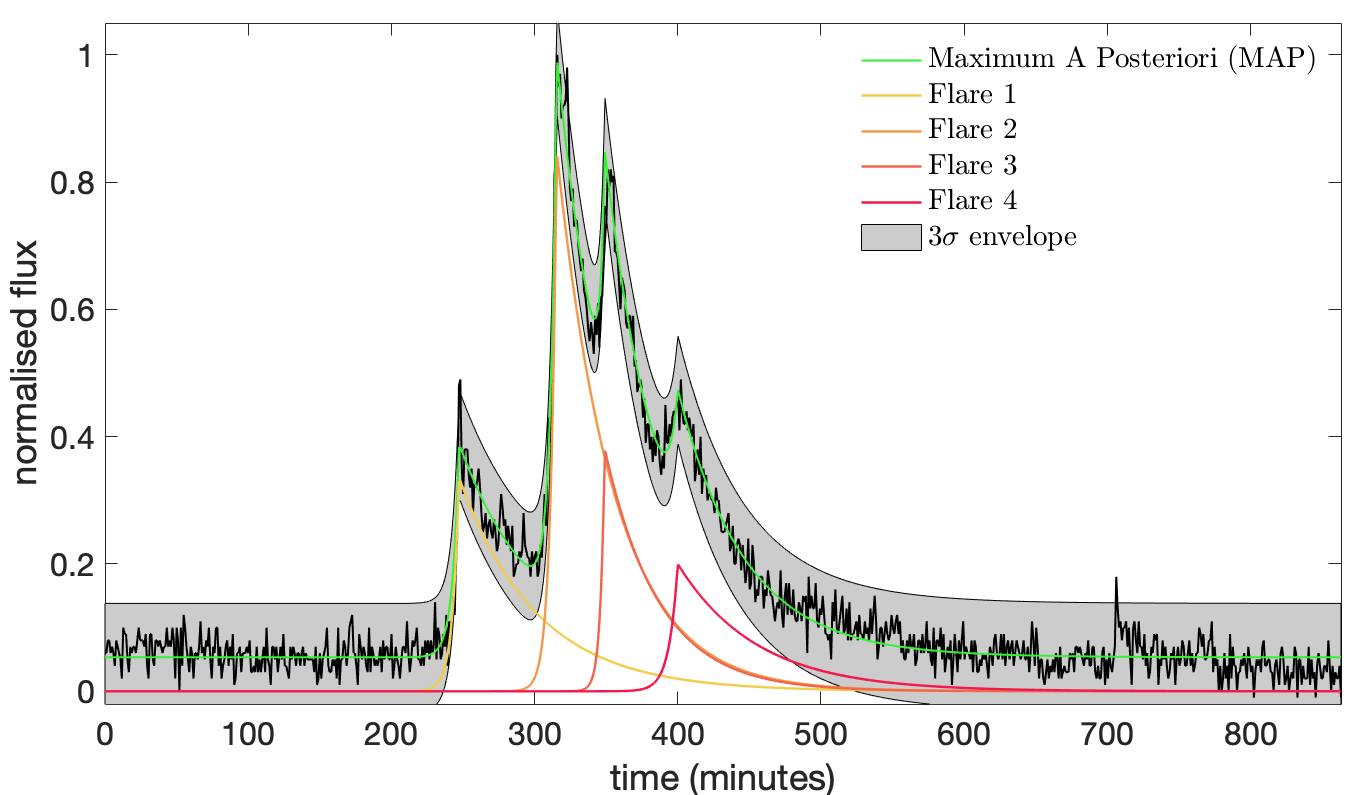}
    \includegraphics[width=\columnwidth]{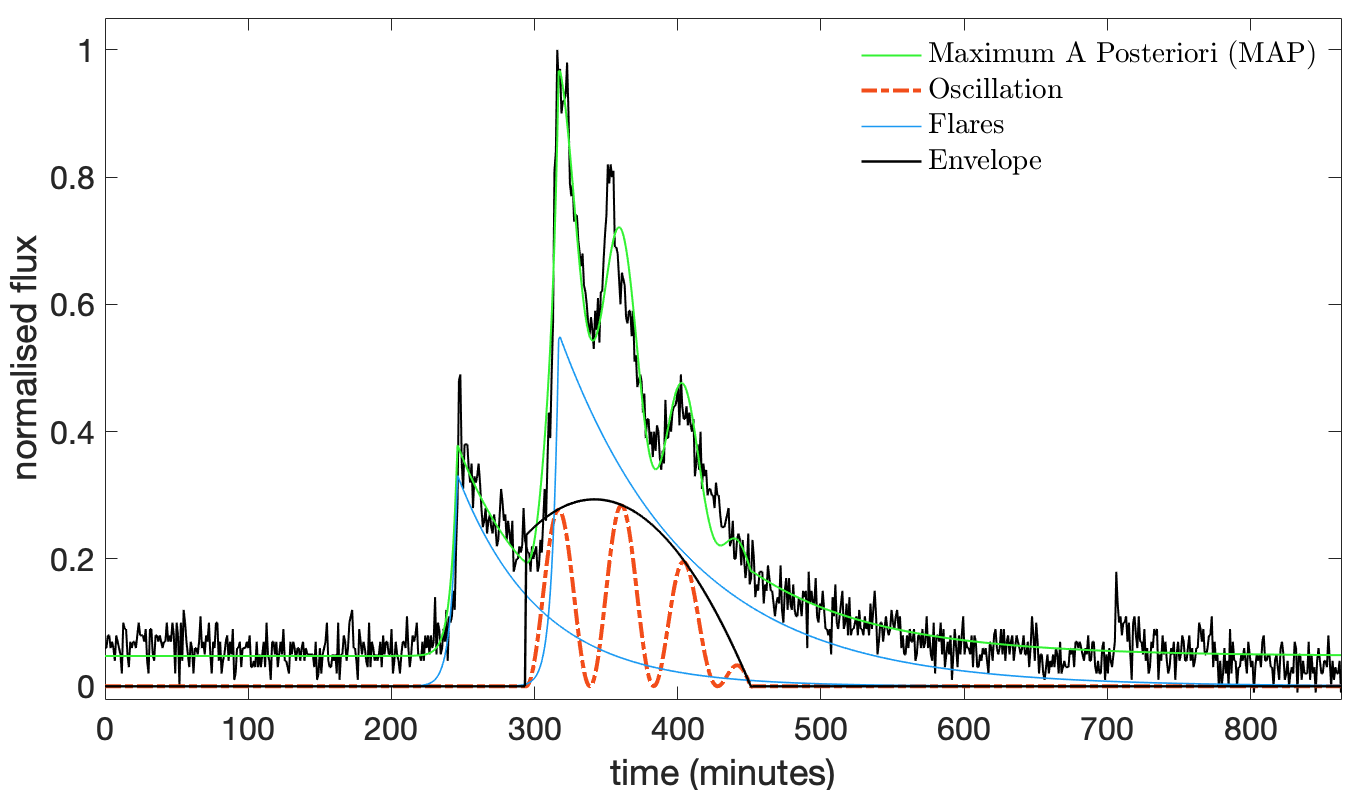}
   \caption{Maximum a posteriori (MAP) from ATAIS for the multi-flare model with four flares (left) 
   and the two-flares model with oscillatory signal triggered during the second flare and a spline
   envelope (right).}
   \label{fig:Kepler}
\end{figure*}

\section{Summary and conclusions}
\label{sec:conclusions}

In this work, we present {an implementation of the adaptive importance sampling method with adaptation of the target (ATAIS) developed in \citet{Martino2021}.} The method accepts different adaptation schemes and includes weight clipping to avoid the so-called \emph{weight {degeneracy}} {problem of importance sampling}. We {remark} that the use of weight clipping is not mandatory but it is recommended for high-dimensional problems to avoid the weight collapse, i.e., that only one {significant sample} is used at each iteration. The main advantage of ATAIS with respect to MCMC methods is that the {model evidence} can be determined {directly} from the importance weights. This makes the comparison between {different models (with the same data)} easier. Compared to other adaptive importance sampling (AIS) methods, ATAIS includes a {target adaptation scheme that uses the mean square error (MSE) in the current iteration to modify the variance of the likelihood function.} With this optimisation scheme, the intensity of the noise is inferred, together with the model parameters. This noise includes not only observational errors but that of model selection/truncation.  

{The performance of the ATAIS algorithm is tested against simulated {and real data}. First, we demonstrate the capability of ATAIS to accurately compute the marginal likelihood within a model selection problem in Section~\ref{sec:ModelSelection}. The simulation is from a sinusoid with very low signal-to-noise ratio. The two models confronted are the sinusoid model and a pure noise model. ATAIS results are compared with those of a numerical integration of the marginal likelihood and with a novel nested sampling method. ATAIS is able to accurately reproduce the marginal likelihood determined with the numerical integration for both models. {Its performance is similar to that of nested sampling schemes}. We then use ATAIS for a bayesian inference problem for higher dimension models. In this test, we simulate the light curve for a single loop flare with an exponentially damped oscillation.} {In addition, we use our algorithm to analyse real data from a flare detected with the space mission \emph{Kepler}. Our results are compared with those obtained by \citet{Pascoe2020} with a combination of MCMC and IS methods. The results are discussed in Section~\ref{sec:inference}. The method is able to discriminate between different models including oscillations and multi-flare models by comparing their model evidences. The latter are outcomes from the method. For the case of real data, our results are similar to those obtained by \citet{Pascoe2020}.}
ATAIS is a powerful tool for Bayesian inference problems. It is meant for model selection problems where the number of {unknown parameters varies across the candidate models. For example, in the comparison of models with different numbers of planets or the distinct models of ignition of flaring events in solar type stars as those analysed in Section \ref{sec:inference}.} ATAIS is also appealing if the model noise is not known a priori. This method includes an optimisation scheme to determine the noise variance. No tempering schemes, of the type commonly used in MCMC methods, are needed.} {Neither is inferring the noise variance as yet another parameter, thereby increasing the dimension of the model}.

\section*{Acknowledgements}

This work was supported by the {Office of Naval Research} (N00014-19-1-2226), Spanish Ministry of Science and Innovation (CLARA; RTI2018-099655-B-I00) and Regional Ministry of Education and Research for the Community of Madrid (PRACTICO; Y2018/TCS-4705).
{This paper includes data collected by the Kepler mission and obtained from the MAST data archive at the Space Telescope Science Institute (STScI). Funding for the Kepler mission is provided by the NASA Science Mission Directorate. STScI is operated by the Association of Universities for Research in Astronomy, Inc., under NASA contract NAS 5–26555. The authors acknowledges fruitful discussion with the referee to improve the manuscript.}

\section*{Data Availability}

The simulated data underlying this article will be shared on reasonable request to the corresponding author. The algorithms developed for this work are coded in Matlab\footnote{https://www.mathworks.com/products/matlab.html}. They will be shared on reasonable request to the corresponding author. 




\bibliographystyle{mnras}
\bibliography{ATAIS} 








\bsp	
\label{lastpage}
\end{document}